\documentclass{article}
\usepackage{spconf}

\usepackage{subfigure}
\usepackage[utf8]{inputenc}
\usepackage{bm, amssymb, amsfonts, amsmath, amsthm, shortcuts_OPT}
\usepackage{bbm}
\usepackage{stmaryrd}
\usepackage{leftidx}
\usepackage{algorithm,algorithmic}
\usepackage{mathrsfs, enumitem}
\usepackage{cite}

\usepackage{graphicx} 
\usepackage{epstopdf}

\usepackage{tikz}
\usepackage{pgfplots}
\usetikzlibrary{matrix}
\usepackage{filecontents}
\usetikzlibrary{calc}
\usepgfplotslibrary{groupplots,dateplot}
\usetikzlibrary{patterns,shapes.arrows}
\pgfplotsset{compat=newest}

\usepackage[colorlinks=true,
linkcolor=green!25!black,
urlcolor=blue,
citecolor=blue]{hyperref}       
\usepackage{cleveref}

\usepgfplotslibrary{fillbetween}
\usepackage{stfloats}

\usepackage{graphicx}
\usepackage{algorithm,algorithmic}
\usepackage[utf8]{inputenc}
\usepackage[english]{babel}

\newtheorem{Prop}{Proposition}

\newtheorem{Corollary}{Corollary}

\newtheorem{Exa}{Example}
\newtheorem{assumption}{H\!\!}

\ninept

\title{Product Graph Learning from Multi-attribute Graph Signals with Inter-layer Coupling}
\name{Chenyue Zhang, Yiran He, Hoi-To Wai\thanks{This work is supported in part by CUHK Direct Grant \#4055135. Emails: \texttt{\{cyzhang,yrhe,htwai\}@se.cuhk.edu.hk}}}
\address{Department of SEEM, The Chinese University of Hong Kong, Shatin, Hong Kong SAR of China}

\begin{document}

\maketitle

\begin{abstract}
This paper considers learning a product graph from multi-attribute graph signals. Our work is motivated by the widespread presence of multilayer networks that feature interactions \emph{within} and \emph{across} graph layers. Focusing on a product graph setting with homogeneous layers, we propose a bivariate polynomial graph filter model. We then consider the topology inference problems thru adapting existing spectral methods. We propose two solutions for the required spectral estimation step: a simplified solution via unfolding the multi-attribute data into matrices, and an exact solution via nearest Kronecker product decomposition (NKD). Interestingly, we show that strong inter-layer coupling can degrade the performance of the unfolding solution while the NKD solution is robust to inter-layer coupling effects. Numerical experiments show efficacy of our methods. 
\end{abstract}
\begin{keywords}
graph signal processing, product graph learning, multi-attribute graph signals, network inference
\end{keywords}

\section{Introduction}\label{sec:intro}\vspace{-.2cm}
In recent years, there has been a growing trend in data science to develop tools for learning and making inference from signals or data observed on networks. The latter is also known as graph signals which form the subject of investigation in the emerging field of graph signal processing (GSP).
Through modeling real-world networks as graphs and encoding the network data as filtered graph signals, an emerging trend is to develop tools for learning the latent graph topology from these network data; see \cite{dong2019learning, mateos2019connecting} and the references therein. 
These tools have widespread applications in the studies of social, financial, and biology networks \cite{newman2018networks}. 

Previous works on topology learning with GSP models have focused on mono-layer networks that consist of a single network ‘layer’, and the tools developed are applicable to single-way data only. However, this may not yield a faithful model for many complex systems since networks and graphs do not simply live in isolation. Instead, many networked systems are better described by multi-layer networks featuring {coupling} interactions \emph{within} and \emph{across} network layers. Examples include: opinion dynamics on correlated topics \cite{parsegov2016novel}, multi-dimensional diffusion \cite{gomez2013diffusion}, protein-protein interactions \cite{prvzulj2011protein}, animal networks \cite{finn2019use}, and relations in image pixels \cite{shi2000normalized}, etc. Observations made on these complex systems are usually generated from two or more coupled networks, and they give rise to multi-attribute observations on nodes, i.e., multi-way graph signals. Naturally, the graph structure embedded in these network data shall be treated using a multilayer graph model as they can not be captured by simple individual networks or flattened networks without structure. 

This paper is motivated by the above to develop an inter-layer coupling aware GSP framework for learning an accurate graph structure from multi-way data. As a special case, we focus on modeling a {product graph} \cite{hammack2011handbook} with homogeneous graph layers. We also develop the corresponding inference algorithms.
In light of this, we contribute to both modeling and algorithm aspects in GSP with:
\begin{itemize}[leftmargin=*, noitemsep]
    \item We adopt a general product graph filter for modeling multi-attribute (a.k.a.~multi-way) graph signals. The model uses a bivariate polynomial graph filter to describe interactions over the \emph{physical} and \emph{coupling} graphs. We show that it encompasses a number of common dynamics on multilayer networks. 
    \item We develop inference algorithms for topology reconstruction and blind centrality detection in learning the product graph model. Our development relies on an observation that the eigenvectors of graph signals' covariance can be written as Kronecker product of respective eigenvectors of coupling and physical graphs. Leveraging this observation, we suggest spectral methods by proposing two solutions based on the layer/node-wise unfolding and nearest Kronecker product decomposition (NKD). We compare the two solutions by analyzing the effects of inter-layer coupling and show that NKD yields an exact solution under milder assumptions.  
\end{itemize}
Lastly, we present numerical experiments on synthetic and real data to corroborate with our analysis. The experiments highlight the effects of inter-layer coupling strengths on topology inference.\vspace{.1cm}

\begin{figure}[t]
\centering
\resizebox{.9\linewidth}{!}{\sf \includegraphics{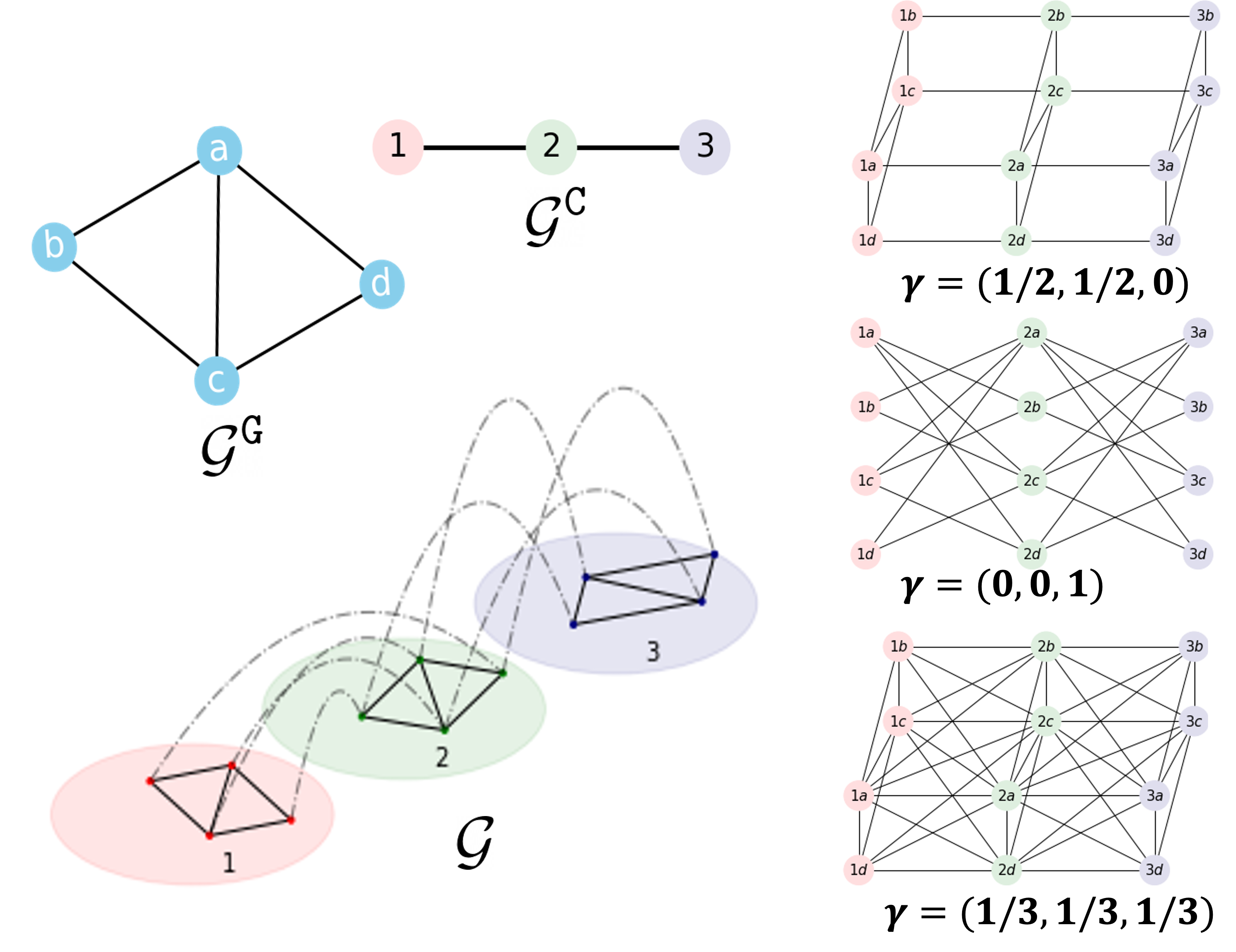}}\vspace{-.3cm}
\caption{(Left) Physical graph $\mathcal{G}^{\tt G}$ and coupling graph $\mathcal{G}^{\tt C}$, together with the overall product graph ${\cal G}$. (Right) Interaction graph ${\cal G}^{\tt I}$ with different set of parameters $\bm{\gamma}$. }\vspace{-.5cm}
\label{fig:cp}
\end{figure}

\noindent \textbf{Related Works.} 
The closest works to ours are \cite{kadambari2021product,kalaitzis2013bigraphical} which studied product graph learning from multi-way graph signals focusing on a smoothness condition defined with the Cartesian product graph. Such conditions restricted the application of these techniques to a special form of coupling mechanism in the multi-way data, which are relaxed by our general product graph filter model.
Besides graph learning, the works \cite{loukas2019stationary,grassi2017time} proposed a time-vertex GSP framework modeling the temporal evolution of graph signals. These models can be regarded as a special case of ours where the coupling graph is fixed as a known path graph. 
Moreover, \cite{sandryhaila2014big, stanley2020multiway,zhang2021introducing} studied models for the generic multi-way GSP with possibly heterogeneous graph layers. While the graph model is considerably more generic, these works focused on simple graph filters, which do not account for a number of inter-layer coupling mechanisms. Lastly, we note that \cite{butler2022convolutional} studied an alternative multigraph GSP model without multi-way data. 
To our best knowledge, this work introduces the first product graph model with general inter-layer coupling accompanied by a suite of topology inference algorithms.\vspace{-.1cm}


\section{Multi-attribute Graph Signals}\vspace{-.2cm}
This section introduces a generative model for multi-attribute graph signals inspired by network dynamic processes on multiplex graphs with homogeneous graph layers. In particular, our development involves a product graph model \cite{hammack2011handbook} and a general multi-dimensional graph signal/filter model based on the former.\vspace{.1cm}

\noindent \textbf{Product Graph Model.} 
The center of our study is a \emph{product graph} ${\cal G} = ({\cal G}^{\tt{C}}, {\cal G}^{\tt{G}})$ formed by two undirected graphs, denoted respectively as $\mathcal{G}^{\tt{C}} =(\mathcal{V}^{\tt{C}}, \mathcal{E}^{\tt{C}},{\bm A}^{\tt{C}})$,  $\mathcal{G}^{\tt{G}}=(\mathcal{V}^{\tt{G}}, \mathcal{E}^{\tt{G}},{\bm A}^{\tt{G}})$ with $|\mathcal{V}^{\tt{C}}|=M$, $|\mathcal{V}^{\tt{G}}|=N$. Note that ${\bm A}^{\tt{C}}$, ${\bm A}^{\tt{G}}$ are weighted adjacency matrices associated with the respective graphs of the edge sets $\mathcal{E}^{\tt{C}} \subseteq \mathcal{V}^{\tt{C}} \times \mathcal{V}^{\tt{C}}$, $\mathcal{E}^{\tt{G}} \subseteq \mathcal{V}^{\tt{G}} \times \mathcal{V}^{\tt{G}}$. We adopt the interpretation of ${\cal G}$ as a \emph{multi-layer graph} such that each node in ${\cal V}^{\tt{G}}$ corresponds to a physical entity, e.g., individual on a social network, etc., and each node in ${\cal V}^{\tt{C}}$ corresponds to a layer/attribute of these physical entities. In this way, ${\cal G}^{\tt{G}}$ will be referred as the \emph{physical graph}, while ${\cal G}^{\tt{C}}$ is the \emph{coupling graph} between layers/attributes; see Fig.~\ref{fig:cp} (left). 

Compared to conventional single graph layer models, an important feature of ${\cal G}$ is that it explicitly models \emph{inter-layer coupling} where all $N$ nodes co-exist in the $M$ interacting layers. It is then instrumental to define the \emph{interaction graph} ${\cal G}^{\tt I}$ with the node set ${\cal V}^{\tt I} = {\cal V}^{\tt{C}} \times {\cal V}^{\tt{G}}$. Notice that $|{\cal V}^{\tt I}| = NM$ and the edge set of ${\cal G}^{\tt I}$ can then be encoded into the adjacency matrix:\vspace{-.1cm}
\beq \label{eq:gene_graph}
\A^{\tt I}=\gamma_1\I\otimes \A^{\tt{G}} + \gamma_2\A^{\tt{C}} \otimes\I + \gamma_3\A^{\tt{C}} \otimes\A^{\tt{G}},\vspace{-.1cm}
\eeq
where $\bm{\gamma} = (\gamma_1,\gamma_2,\gamma_3) \geq 0$ satisfying $\sum_{i=1}^3 \gamma_i = 1$ are the \emph{coupling parameters} of the generalized product graph, and $\otimes$ denotes Kronecker product. 
Different combinations of $\bm{\gamma}$ can lead to different interaction patterns among the nodes in ${\cal V}$ \cite{tran2018generalized}. For example, with $\bm{\gamma} = (\frac{1}{2},\frac{1}{2},0)$, $\A^{\tt I}$ describes the Cartesian product graph and the matrix is also known as the supra-adjacency matrix; with $\bm{\gamma} = (0,0,1)$, $\A^{\tt I}$ describes the Kronecker product graph; with $\bm{\gamma} = (\frac{1}{3}, \frac{1}{3}, \frac{1}{3})$, $\A^{\tt I}$ describes the strong product graph; see Fig.~\ref{fig:cp} (right).\vspace{.2cm}


\noindent \textbf{Multi-attribute Graph Signals.}
We concentrate on modeling multi-attribute graph signals living on ${\cal G}^{\tt I}$ where the \emph{graph filter} is induced by the graph factors ${\bm A}^{\tt C}, \A^{\tt G}$ and takes an excitation signal as its input. For example, the filter can be a function of ${\bm A}^{\tt I}$. 
We consider a general formulation of multi-layer graph filter \cite{natali2020forecasting}:\vspace{-.1cm}
\beq \textstyle \label{eq:gf}
{\cal H}( \A^{\tt C}, \A^{\tt G} ) = \sum_{i=0}^{T_{\tt G}} \sum_{j=0}^{T_{\tt C}} h_{ij} (\A^{\tt{C}})^j \otimes (\A^{\tt{G}})^i,\vspace{-.1cm}
\eeq 
where $h_{ij} \in \RR$ are the filter coefficients and $T_{\tt G}, T_{\tt C} \in \ZZ_+ \cup \{ \infty \}$ are the orders of the \emph{bivariate filter polynomial}. The multi-attribute (a.k.a.~multi-way) graph signals are then modeled as:\vspace{-.15cm}
\beq \label{eq:gsp}
{\bm y}^{(s)} = {\cal H}( \A^{\tt C}, \A^{\tt G} ) {\bm x}^{(s)} + {\bm w}^{(s)},\vspace{-.2cm}
\eeq 
where $s \in \NN$ denotes the sample index, ${\bm x}^{(s)}, {\bm w}^{(s)} \in \RR^{NM}$ are the excitation signals and observation noise, respectively. Note that ${\bm y}  = ( {\bm y}^{(1)} ; \cdots ; {\bm y}^{(M)} ) \in \RR^{NM}$ whose $m$th block, ${\bm y}^{(m)}$, corresponds to observations on the nodes $\{ (m,i) \in {\cal V}^{\tt I} : i \in {\cal V}^{\tt G} \}$ and the graph signals have been organized in a layer-by-layer fashion. 

It is obvious that for all coupling parameters $\bm{\gamma}$, any polynomial of ${\bm A}^{\tt I}$ can be written as \eqref{eq:gf}. 
Notably, it is common \cite{gomez2013diffusion} to use the supra-adjacency matrix, ${\bm A}^{\tt I}$ with $\bm{\gamma} = ( \frac{1}{2}, \frac{1}{2}, 0 )$, to describe interactions in a multi-layer graph, and subsequently consider the polynomial of ${\bm A}^{\tt I}$ as the graph filter. However, this is not sufficient to describe certain interactions on ${\cal G}$, e.g., a polynomial of ${\bm A}^{\tt C} \otimes {\bm A}^{\tt G}$ cannot be expressed\footnote{Note that the issue boils down to the non-existence of a polynomial $\tilde{h}(\cdot)$ such that ${h}(ab) = \tilde{h}(a+b)$ for all $h(\cdot), a,b$.} as a polynomial of ${\bm A}^{\tt I}$ with $\bm{\gamma} = ( \frac{1}{2}, \frac{1}{2}, 0 )$. 

We conclude this section by showcasing two example dynamic processes to illustrate the generality of our graph filter model:\vspace{-.1cm}
\begin{Exa} Consider the Friedkin-Johnsen multi-dimensional opinion dynamics \cite{friedkin1990social, parsegov2016novel}. At time $t \geq 0$ and the $s$th discussion, the multi-dimensional opinions for  agents in ${\cal V}^{\tt G}$ evolve as \vspace{-.15cm}
\begin{equation} \label{eq:mul_degroot_dyn}
{\bm x}(t+1) = ( {\bm A}^{\tt C} \otimes {\bm A}^{\tt G} ) {\bm x}(t) + {\bm x}^{(s)}\vspace{-.15cm}
\end{equation}
such that the $m$th block of ${\bm x}(t)$ represents the opinions of $N$ agents on the $m$th topic. The weighted adjacency matrices ${\bm A}^{\tt C}, {\bm A}^{\tt G}$ represent the logical dependencies between the $M$ topics, and the mutual trusts between the $N$ agents, respectively. They are properly scaled such that the vector $( {\bm I}_{NM} - {\bm A}^{\tt C} \otimes {\bm A}^{\tt G} ){\bf 1} > {\bm 0}$ represents the self trusts of each agent on the topic. Moreover, ${\bm x}^{(s)}$ is the initial belief of the agents. Under the above premises, we have \vspace{-.15cm}
\beq 
{\bm y}^{(s)} = \lim_{t \to \infty} {\bm x}(t) = ({\bm I}_{NM} - {\bm A}^{\tt C} \otimes {\bm A}^{\tt G} )^{-1} {\bm x}^{(s)} , \vspace{-.2cm}
\eeq 
where we note that $({\bm I}_{NM} - {\bm A}^{\tt C} \otimes {\bm A}^{\tt G} )^{-1}$ is a special case of \eqref{eq:gf} since the latter is a function of $\A^{\tt I}$ with $\bm{\gamma} = (0,0,1)$.
\end{Exa}
\begin{Exa}
The diffusion process in \cite{gomez2013diffusion} describes dynamics such as social contact, epidemic, etc., on a multi-layer graph. In particular, with the excitation ${\bm x}^{(s)}$, the states of nodes at time $t$ evolve as:\vspace{-.1cm}
\begin{equation} \textstyle 
\frac{ d {\bm x}(t) }{dt} = - {\bm x}(t) + \big( \A^{\tt C} \otimes {\bm I}_N + {\bm I}_M \otimes \A^{\tt G} \big) {\bm x}(t) + {\bm x}^{(s)}.\vspace{-.1cm}
\end{equation}
With properly scaled $\A^{\tt C}, \A^{\tt G}$, the above has a unique equilibrium: \vspace{-.15cm}
\begin{equation}
{\bm y}^{(s)} = ( {\bm I}_{NM} - \A^{\tt C} \otimes {\bm I}_N - {\bm I}_M \otimes \A^{\tt G} )^{-1} \x^{(s)}.\vspace{-.1cm}
\end{equation}
where we note that $( {\bm I}_{NM} - \A^{\tt C} \otimes {\bm I}_N - {\bm I}_M \otimes \A^{\tt G} )^{-1}$ is a special case of \eqref{eq:gf} since the latter is a function of $\A^{\tt I}$ with $\bm{\gamma} = (\frac{1}{2},\frac{1}{2},0)$.\vspace{-.1cm}
\end{Exa}
\noindent Graph filter models in the form of \eqref{eq:gf} may also be found in other data scenarios such as the graph causal processes \cite{mei2016signal}. These models may be based on other types of interaction graphs ${\cal G}^{\tt I}$ such as strong product, but they can nevertheless be covered by \eqref{eq:gf}.\vspace{-.2cm}

\section{Product Graph Learning}\vspace{-.2cm}
We aim to infer the product graph model through observing the multi-attribute graph signals in \eqref{eq:gsp}. We concentrate on learning the coupling and physical graphs ${\bm A}^{\tt C}, {\bm A}^{\tt G}$. Our idea is to apply spectral methods with reliable performance regardless of the graph filter \eqref{eq:gf}.

To fix ideas, we denote the eigenvalue decompositions (EVDs) for the adjacency matrices by:\vspace{-.1cm}
\beq 
\A^{\tt C} = \V^{\tt C} \bm{\Lambda}^{\tt C} (\V^{\tt C})^\top, \A^{\tt G} = \V^{\tt G} \bm{\Lambda}^{\tt G} (\V^{\tt G})^\top,\vspace{-.1cm}
\eeq 
where $\V^{\tt C}, \V^{\tt G}$ are orthogonal and $\bm{\Lambda}^{\tt C}, \bm{\Lambda}^{\tt G}$ are diagonal matrices. Without loss of generality, the eigenvalues are sorted in decreasing order. For example, the $i$th column vector of ${\bm V}^{\tt G}$, denoted ${\bm v}_i^{\tt G}$, corresponds to the $i$th largest eigenvalue ($\lambda_i^{\tt G}$) in ${\bm A}^{\tt G}$. We assume that:\vspace{-.1cm}
\begin{assumption} \label{ass:dis}
For any $i=1,\ldots,N$, $j=1,\ldots,M$, the magnitudes of frequency response $|h(\lambda_j^{\tt C},\lambda_i^{\tt G})|$ have distinct values, where we defined $h(\lambda^{\tt C}, \lambda^{\tt G}) := \sum_{i=0}^{T_{\tt G}}\sum_{j=0}^{T_{\tt C}} h_{ij} (\lambda^{\tt C})^j (\lambda^{\tt G})^i$ according to \eqref{eq:gf}.\vspace{-.1cm}
\end{assumption}
\noindent The above assumption holds for cases when the graph filter is a function of ${\bm A}^{\tt I}$ where ${\bm A}^{\tt C}, {\bm A}^{\tt G}$ have distinct eigenvalues. 
Consequently, the EVD of the graph filter \eqref{eq:gf} is derived as (cf.~\cite{sandryhaila2014big}):\vspace{-.1cm}
\beq 
{\cal H}( \A^{\tt C}, \A^{\tt G} ) = ( \V^{\tt C} \otimes \V^{\tt G} ) {\cal H}( \bm{\Lambda}^{\tt C}, \bm{\Lambda}^{\tt G} )  ( \V^{\tt C} \otimes \V^{\tt G} )^\top , \vspace{-.1cm}
\eeq 
such that ${\cal H}( \bm{\Lambda}^{\tt C}, \bm{\Lambda}^{\tt G} )$ is a diagonal matrix. Under standard white noise conditions for ${\bm x}^{(s)}, {\bm w}^{(s)}$, i.e., both are zero-mean and satisfy $\EE[{\bm x}^{(s)} ({\bm x}^{(s)})^\top] = {\bm I}$, $\EE[{\bm w}^{(s)} ({\bm w}^{(s)})^\top] = \sigma^2 {\bm I}$, the graph signal covariance ${\bm C}_y = \EE[ {\bm y}^{(s)} ({\bm y}^{(s)})^\top ]$ can be derived as:\vspace{-.1cm}
\beq \label{eq:covy}
{\bm C}_y = 
( \V^{\tt C} \otimes \V^{\tt G} ) |{\cal H}( \bm{\Lambda}^{\tt C}, \bm{\Lambda}^{\tt G} )|^2  ( \V^{\tt C} \otimes \V^{\tt G} )^\top + \sigma^2 {\bm I}.\vspace{-.1cm}
\eeq 
Eq.~\eqref{eq:covy} makes an important observation about the covariance matrix ${\bm C}_y$. When $\sigma = 0$, the eigenvectors of ${\bm C}_y$ are given by the columns of ${\bm V}^{\tt C} \otimes {\bm V}^{\tt G}$. We will demonstrate in \S\ref{sec:estV} that ${\bm V}^{\tt C}, {\bm V}^{\tt G}$ can be retrieved by decomposing the eigenvectors of ${\bm C}_y$. 

Suppose for now that 
the matrices $\widehat{\bm V}^{\tt C}, \widehat{\bm V}^{\tt G}$ with possibly permuted columns of $\V^{\tt C}, \V^{\tt G}$ are given, we can recover the graph topology and/or detect central nodes in ${\cal G}^{\tt C}, {\cal G}^{\tt G}$ through adapting the spectral methods developed in several existing works. In particular:\vspace{.1cm}

\noindent \textbf{P1. Topology Reconstruction.} It is shown in \cite{segarra2017network} that graph topologies can be reconstructed from the spectral template of graph shift operator (GSO). For example, the following problem recovers ${\bm A}^{\tt G}$:\vspace{-.1cm}
\begin{align} 
 \textstyle \min_{ \bm{\lambda}^{\tt G} , \widehat{\bm A}^{\tt G} }~ & \textstyle \|{\rm vec}( \widehat{\bm A}^{\tt G})\|_1 + \frac{\rho}{2} \| \widehat{\bm A}^{\tt G} - \widehat\V^{\tt G} {\rm Diag}(\bm{\lambda}^{\tt G}) (\widehat\V^{\tt G})^\top \|_F^2 \notag \\[.0cm]
\text{s.t.}~ & |{\rm diag}(\widehat{\bm A}^{\tt G})| \leq \epsilon {\bf 1}, \widehat{\bm A}^{\tt G} {\bf 1} \geq {\bf 1}, \label{eq:topo_recon} \\[-.6cm] \notag
\end{align} 
where $\rho > 0, \epsilon > 0$ are regularization parameters, and ${\rm vec}(\cdot)$ is the vectorization operator. 
Similar formulation can be applied for $\A^{\tt C}$. \vspace{.1cm}

\noindent \textbf{P2. Centrality Estimation.} 
It is observed \cite{roddenberry2021blind, he2022detecting, he2021identifying} that eigen-centrality vector can be inferred from graph signals under mild conditions. 
Their idea can be adapted for product graphs by noting ${\bm v}_1^{\tt C} \otimes {\bm v}_1^{\tt G}$ is the only positive eigenvector in ${\bm C}_y$. Under the premise that the latter can be decomposed (cf.~\S\ref{sec:estV}), we propose:
\vspace{-.1cm}
\beq \textstyle \label{eq:centrali}
\widehat{\bm c}^{\tt G} = \widehat{\bm v}^{\tt G}_c,~\widehat{\bm c}^{\tt C} = \widehat{\bm v}^{\tt C}_c~~\text{where}~~\widehat{\bm v}_{i^\star} = \widehat{\bm v}^{\tt C}_c \otimes \widehat{\bm v}^{\tt G}_c\vspace{-.1cm}
\eeq 
and $i^\star = \argmin_i {\sf P}( \widehat{\bm v}_i )$ with
${\sf P}( {\bm x} ) := \min\{ \| {\bm x} - ({\bm x})^+ \|, \| {\bm x} + (-{\bm x})^+ \| \}$, $({\bm x})^+ := \max\{ {\bf 0}, {\bm x} \}$, and ${\sf P}({\bm x}) = 0$ iff ${\bm x}$ is all-positive or all-negative. Notice $\widehat{\bm v}_i$ denotes the $i$th eigenvector of ${\bm C}_y$.\vspace{-.2cm}


\subsection{Recovering ${\bf V}^{\tt C}, {\bf V}^{\tt G}$ from ${\bf C}_y$} \label{sec:estV} \vspace{-.1cm}
Our remaining task is to recover the eigenvectors in ${\bm V}^{\tt C}, {\bm V}^{\tt G}$ \emph{individually} from observed data ${\bm y}^{(s)}$. For simplicity, we consider the same conditions leading to \eqref{eq:covy} with $\sigma = 0$ and propose two solutions.\vspace{.1cm}

\noindent \textbf{Exact Solution by NKD.} Observe that the diagonal entries in $|{\cal H}(\bm{\Lambda}^{\tt C}, \bm{\Lambda}^{\tt G})|^2$ of \eqref{eq:covy} may not be sorted in descending order. Applying EVD on ${\bm C}_y$ thus produces the eigenvector matrix:
\vspace{-.1cm}
\beq \label{eq:vhat_cov}
\widehat{\bm V} = (\V^{\tt C} \otimes \V^{\tt G}) \bm{\Pi} = \big[ \cdots~{\bm v}_{\pi^{\tt C}(i)}^{\tt C} \otimes {\bm v}_{\pi^{\tt G}(i)}^{\tt G} \cdots \big],\vspace{-.1cm}
\eeq 
where $\bm{\Pi}$ is a permutation matrix that orders the columns of $\V^{\tt C} \otimes \V^{\tt G}$ according to the magnitude of frequency response. For any $i \in \{1,\ldots,NM\}$, $\pi^{\tt C}(i), \pi^{\tt G}(i)$ are indices of eigenvectors for $\V^{\tt C}, \V^{\tt G}$ of the $i$th highest frequency response in $|{\cal H}(\bm{\Lambda}^{\tt C}, \bm{\Lambda}^{\tt G})|^2$, i.e., $\widehat{\bm v}_i$.

The structure illustrated in \eqref{eq:vhat_cov} indicates that every column vectors of $\widehat{\V}$ can be written as a Kronecker product. 
To this end, we observe the classical result adapted from \cite[Theorem 2.1]{van1993approximation}.\vspace{-.1cm}
\begin{Prop} \label{prop:van}
Consider ${\bm M} \in \RR^{MN \times PQ}$, ${\bm A} \in \RR^{M \times P}$, ${\bm B} \in \RR^{N \times Q}$. 
If ${\bm M} = {\bm A} \otimes {\bm B}$, then ${\cal R}({\bm M}) = {\rm vec}( {\bm A} ) {\rm vec}( {\bm B} )^\top$, where \vspace{-.1cm}
\[
{\cal R}( {\bm M} ) = \left[ \begin{array}{c} 
{\bm M}_1 \\ \vdots \\ {\bm M}_P 
\end{array} \right],~
{\bm M}_i = \left[ 
\begin{array}{c} 
{\rm vec}({\bm M}_{1,i})^\top \\ \vdots \\ {\rm vec}({\bm M}_{M,i})^\top
\end{array} \right],~i=1,\ldots,P.\vspace{-.1cm}
\]
Note ${\bm M}$ has been partitioned into $N \times Q$ blocks and ${\bm M}_{i,j}$ is the $(i,j)$th block. 
Subsequently, ${\bm A}, {\bm B}$ can be uniquely  recovered (up to a scalar factor) through a suitable decomposition of ${\cal R}({\bm M})$.\vspace{-.1cm}
\end{Prop}
\noindent When ${\bm M} \neq {\bm A} \otimes {\bm B}$, e.g., with noisy observations, \cite{van1993approximation} consider the \emph{nearest Kronecker product decomposition} (NKD) problem:\vspace{-.1cm}
\beq \textstyle \label{eq:nkp}
\begin{array}{rl}
\min_{ {\bm A}, {\bm B} , \alpha } & \| {\cal R}( {\bm M} ) - \alpha \, {\rm vec}( {\bm A} ) {\rm vec}( {\bm B} )^\top \|_F^2 \\
\text{s.t.} & \| {\rm vec}( {\bm A} ) \| = 1,~\| {\rm vec}( {\bm B} ) \| = 1,\vspace{-.1cm}
\end{array}
\eeq 
which can be solved by finding the top singular vectors of ${\cal R}( {\bm M} )$.

Proposition~\ref{prop:van} shows that ${\bm v}_{\pi^{\tt C}(i)}^{\tt C}, {\bm v}_{\pi^{\tt G}(i)}^{\tt G}$ can be recovered by applying NKD on ${\cal R}( \widehat{\bm v}_i )$ from \eqref{eq:vhat_cov}. Consequently, collecting the NKD outputs on ${\cal R}( \widehat{\bm v}_i )$ for all $i$ forms two matrices whose columns are permuted and repeated copies of the columns of $\V^{\tt C}, \V^{\tt G}$. This suggests applying the Gram-Schmidt procedure on these matrices [cf.~\eqref{eq:gs}] for estimating $\V^{\tt C}, \V^{\tt G}$; see Algorithm~\ref{algo:prod_learn}. 
The algorithm returns an exact solution under H\ref{ass:dis} and other conditions:
\vspace{-.1cm}
\begin{Corollary}
Under H\ref{ass:dis} and noiseless observations ($\sigma=0$) with $S \to \infty$, Algorithm~\ref{algo:prod_learn} recovers the columns of $\V^{\tt C}, \V^{\tt G}$.\vspace{-.1cm}
\end{Corollary}
\noindent We remark that for {\bf P2}, one may skip step 4 of Algorithm~\ref{algo:prod_learn} to obtain $i^\star = \argmin_i {\sf Pos}( \widehat{\bm v}_i^{\sf noisy} )$ in \eqref{eq:centrali} and thus the pair $( \widehat{\bm v}_{i^\star}^{\tt C}, \widehat{\bm v}_{i^\star}^{\tt G} )$. In this way, H\ref{ass:dis} can be further weakened to guarantee exact recovery.
 
\begin{figure}[t]
\vspace{-.3cm}
\begin{algorithm}[H]
\caption{Learning Product Graph ({\bf P1} and/or {\bf P2})} 
\label{algo:prod_learn}
\begin{algorithmic}[1]
\STATE \textbf{INPUT}: Set of $S$ observed graph signals $\{ {\bm y}^{(s)} \}_{s=1}^S$.
\STATE Evaluate $\CovYhat^S = (1/S)\sum_{s=1}^S {\bm y}^{(s)} ({\bm y}^{(s)})^\top$ and compute its eigenvectors as $\hatV^{\sf noisy} = {\sf EVD}( \CovYhat^S )$.
\STATE For any $i=1,\ldots,NM$, solve the NKD problem for ${\cal R}( \widehat{\bm v}_i^{\sf noisy} )$ [cf.~\eqref{eq:nkp}] to obtain the pair $( \widehat{\bm v}_i^{\tt C}, \widehat{\bm v}_i^{\tt G} )$. 
\STATE Perform Gram-Schmidt (GS) to obtain the orthogonal matrices:\vspace{-.1cm}
\beq \label{eq:gs}
\widehat{\V}^{\tt C} = {\sf GS}( [\widehat{\bm v}_1^{\tt C}, \ldots, \widehat{\bm v}_{NM}^{\tt C}]),~
\widehat{\V}^{\tt G} = {\sf GS}([\widehat{\bm v}_1^{\tt G}, \ldots, \widehat{\bm v}_{NM}^{\tt G}]).\vspace{-.45cm}
\eeq 
\STATE Apply the methods in {\bf P1} for topology reconstruction \eqref{eq:topo_recon} or {\bf P2} for centrality detection \eqref{eq:centrali}.
\end{algorithmic}
\end{algorithm}
\vspace{-3em}
\end{figure}

\vspace{.15cm}
\noindent \textbf{Simplified Solution by Unfolding.} We conclude the section by proposing and analyzing an alternative to NKD in Algorithm~\ref{algo:prod_learn} for estimating $\V^{\tt C}, \V^{\tt G}$.  
The alternative solution is inspired by \cite{sandryhaila2014big,zhang2021graph} through unfolding the multi-attribute graph signals into layer-wise and node-wise matrices. Interestingly, we show that this simplified design can be as effective as NKD, but only when the frequency response of the product graph filter \eqref{eq:gf} is \emph{separable} \cite{grassi2017time}.

\renewcommand{\UrlFont}{\normalsize}

To this end, we denote the unfolding of ${\bm y}^{(s)}$ in \eqref{eq:gsp} as ${\bm Y}^{(s)} = [ {\bm y}_1^{(s)}, \ldots, {\bm y}_M^{(s)} ]$ and consider ${\bm C}_y^{\sf layer} = \EE[ ({\bm Y}^{(s)})^\top {\bm Y}^{(s)} ]$, ${\bm C}_y^{\sf node} = \EE[ {\bm Y}^{(s)} ({\bm Y}^{(s)})^\top ]$. Their resultant eigenvector matrices $\widetilde{\bm V}^{\tt C} = {\sf EVD}( {\bm C}_y^{\sf layer} )$, $\widetilde{\bm V}^{\tt G} = {\sf EVD}( {\bm C}_y^{\sf node} )$ can then be used in lieu of the estimates in step~4 of Algorithm~\ref{algo:prod_learn}. The following proposition analyzes the covariance matrices:\vspace{-.15cm}
\begin{Prop} \label{prop:sep}
Under H\ref{ass:dis}. Assume that the excitation ${\bm x}^{(s)}$ is zero-mean satisfying $\EE[ {\bm x}^{(s)} ( {\bm x}^{(s)} )^\top ] = {\bm I}$ and ${\bm w}^{(s)} = {\bm 0}$. It holds:\vspace{-.2cm}
\beq 
\begin{split} 
{\bm C}_y^{\sf layer} & \textstyle = \sum_{i=1}^M {\bm v}_i^{\tt C} ( {\bm v}_i^{\tt C} )^\top \sum_{j=1}^N | h ( \lambda_i^{\tt C}, \lambda_j^{\tt G} )|^2, \\
{\bm C}_y^{\sf node} & \textstyle = \sum_{j=1}^N {\bm v}_j^{\tt G} ( {\bm v}_j^{\tt G} )^\top \sum_{i=1}^M | h ( \lambda_i^{\tt C}, \lambda_j^{\tt G} )|^2.\\[-.15cm]
\end{split}
\eeq 
\end{Prop}
\noindent The proof can be found in the appendix. 
To gain insight, we focus on node-wise covariance ${\bm C}_y^{\sf node}$ whose eigenvectors $\widetilde{\V}^{\tt G}$ correspond to desired $\V^{\tt G}$ iff the eigenvalues $\sum_{i=1}^M | h ( \lambda_i^{\tt C}, \lambda_j^{\tt G} )|^2$ are distinct. However, this scenario of repeated eigenvalues is common for ${\bm C}_y^{\sf node}$ when $h ( \lambda_i^{\tt C}, \lambda_j^{\tt G} ) = h ( \lambda_i^{\tt C} \lambda_j^{\tt G} )$, i.e., the graph filter is based on interaction graphs with $\bm{\gamma} = (0,0,1)$. For example, when $h ( \lambda_i^{\tt C}, \lambda_j^{\tt G} ) = e^{ \lambda_i^{\tt C} \lambda_j^{\tt G} }$ with the spectrum for ${\cal G}^{\tt C}$ as $\{-1,1,2\}$, then ${\bm C}_y^{\sf node}$ has repeated eigenvalues if $\{ 0, -1.27 \}$ are in the spectrum of ${\cal G}^{\tt G}$.

On the other hand, a sufficient condition for ${\bm C}_y^{\sf node}$ to admit distinct eigenvalues hinges on the \emph{separable filter} property \cite{grassi2017time}. 
Concretely, we need $h(\lambda^{\tt C}, \lambda^{\tt G}) = h^{\tt C}(\lambda^{\tt C}) h^{\tt G}( \lambda^{\tt G})$ for all $\lambda^{\tt C}, \lambda^{\tt G}$. For example, when $h ( \lambda_i^{\tt C}, \lambda_j^{\tt G} ) = e^{ \frac{1}{2}(\lambda_i^{\tt C} + \lambda_j^{\tt G}) }$, the interaction graph is free of the direct Kronecker product. In this case, the eigenvectors of ${\bm C}_y^{\sf layer}, {\bm C}_y^{\sf node}$ serve as good surrogates for $\V^{\tt C}, \V^{\tt G}$, respectively.\vspace{-.1cm}

\vspace{-.2cm}
\section{Numerical Experiments}\vspace{-.2cm} 

\noindent \textbf{Synthetic Data.}
We consider graph filters that are based on ${\bm A}^{\tt I}$ in \eqref{eq:gene_graph} with the parameters $\bm{\gamma} = ( \gamma_1, 2 \gamma_1, 1-3\gamma_1 )$. As $\gamma_1 \downarrow 0$, the interaction graph has strong \emph{inter-layer coupling} of Kronecker product form. 
We fix $\mathcal{G}^{\tt C}$ to be the tree graph in Fig.~\ref{fig:cp} with $M=3$ where ${\bm A}^{\tt C}$ is the unweighted adjacency matrix; while ${\bm A}^{\tt G}$ will be the unweighted adjacency matrix for ${\cal G}^{\tt G}$ to be determined below.\vspace{.1cm}

\noindent {\textbf{Example 1: Topology Reconstruction.}}
This example examines the performance of product graph topology reconstruction ({\bf P1}). We generate ${\cal G}^{\tt G}$ as an Erdos-Renyi graph with connection probability of $p=0.4$. We benchmark the proposed `NKD' based Algorithm~\ref{algo:prod_learn} against `Unfold' which estimates $\widetilde{\V}^{\tt C}, \widetilde{\V}^{\tt G}$ from the layer-wise/node-wise unfolded graph signals; `PGL' refers to the product graph learning method based on smoothness of layer-wise/node-wise unfolded graph signals in \cite{kadambari2021product}; `Flatten' refers to applying \cite{segarra2017network} directly to  ${\bm C}_y$. For the {\tt SpecTemp} problem \eqref{eq:topo_recon} in Algorithm~\ref{algo:prod_learn}, `Unfold', and `Flatten', we take $\rho = 40$, $\epsilon = 10^{-6}$ and solve \eqref{eq:topo_recon} by \texttt{cvx}. 

\begin{figure}[t]
\centering
\resizebox{0.995\linewidth}{!}{\sf \definecolor{mycolor1}{rgb}{0.00000,0.44700,0.74100}%
\definecolor{mycolor2}{rgb}{0.85000,0.32500,0.09800}%
\definecolor{mycolor3}{rgb}{0.49400,0.18400,0.55600}%
\definecolor{mycolor4}{rgb}{0.92900,0.69400,0.12500}%

\begin{tikzpicture}

\begin{groupplot}[group style={group name=myplot,group size=2 by 1}]
\nextgroupplot[%
width=2.5in,
height=2.25in,
xmin=100,
xmax=2000,
xtick = {100,500,1000,1500,2000},
xlabel style={font=\color{white!15!black}},
xlabel={Sample size $S$},
ymin=0,
ymax=1,
x grid style={white!69.0196078431373!black},
xmajorgrids,
y grid style={white!69.0196078431373!black},
ymajorgrids,
ylabel style={font=\color{white!15!black}},
ylabel={$F_{1}$-score},
axis background/.style={fill=white},
legend style={legend cell align=left, align=left, draw=white!15!black}
]
\addplot [very thick, color=mycolor2, mark=o, mark options={solid, mycolor2}]
  table[row sep=crcr]{%
100	0.521551043089051\\
500	0.660710305797389\\
1000	0.695166830141071\\
1500	0.734083677200606\\
2000	0.741092836791656\\
};\label{plots:NKD(a)}

\addplot [very thick, color=mycolor1, mark=o, mark options={solid, mycolor1}]
  table[row sep=crcr]{%
100	0.480229761865191\\
500	0.517961278181142\\
1000	0.531489111932356\\
1500	0.551871406545057\\
2000	0.567520405653698\\
};\label{plots:Sep(a)}

\addplot [color=mycolor3, mark=o, mark options={solid, mycolor3}]
  table[row sep=crcr]{%
100	0.137042252293842\\
500	0.204897987469357\\
1000	0.242939809702312\\
1500	0.250438464458999\\
2000	0.252178459618518\\
};\label{plots:Pgl(a)}

\addplot [color=mycolor4, mark=o, mark options={solid, mycolor4}]
  table[row sep=crcr]{%
100	0.0971220190550182\\
500	0.0954110395738399\\
1000	0.0943004822135386\\
1500	0.0953723507370251\\
2000	0.0955937300487882\\
};\label{plots:Flat(a)}

\addplot [very thick, color=mycolor2, dashed, mark=asterisk, mark options={solid, mycolor2}]
  table[row sep=crcr]{%
100	0.560232988090998\\
500	0.631909212452792\\
1000	0.652896241713461\\
1500	0.652812537557974\\
2000	0.678335998177225\\
};\label{plots:NKD(b)}

\addplot [very thick, color=mycolor1, dashed, mark=asterisk, mark options={solid, mycolor1}]
  table[row sep=crcr]{%
100	0.62984120755054\\
500	0.750502574885727\\
1000	0.775072121674213\\
1500	0.795598353134223\\
2000	0.800359624060028\\
};\label{plots:Sep(b)}

\addplot [color=mycolor3, dashed, mark=asterisk, mark options={solid, mycolor3}]
  table[row sep=crcr]{%
100	0.483183106331304\\
500	0.575296980521579\\
1000	0.592565851258995\\
1500	0.594372667553206\\
2000	0.598095840241754\\
};\label{plots:Pgl(b)}

\addplot [color=mycolor4, mark=asterisk, mark options={solid, mycolor4}]
  table[row sep=crcr]{%
100	0.096162367807108\\
500	0.0961838791772848\\
1000	0.0980815917446714\\
1500	0.096191236683723\\
2000	0.0921184163832996\\
};\label{plots:Flat(b)}
\coordinate (top) at (rel axis cs:0,1);

\nextgroupplot[%
width=2.5in,
height=2.25in,
xmin=5,
xmax=30,
xtick = {5,10,20,30},
xlabel style={font=\color{white!15!black}},
xlabel={Graph Size $N = |{\cal V}^{\tt G}|$},
ymin=0,
ymax=1,
ylabel style={font=\color{white!15!black}},
x grid style={white!69.0196078431373!black},
xmajorgrids,
y grid style={white!69.0196078431373!black},
ymajorgrids,
axis background/.style={fill=white},
legend style={legend cell align=left, align=left, draw=white!15!black}
]
\addplot [very thick, color=mycolor2, mark=o, mark options={solid, mycolor2}]
 table[row sep=crcr]{%
5	0.885834820183023\\
10	0.705617990695929\\
15	0.639581263483261\\
20	0.596132732868606\\
25	0.566283364678718\\
30	0.555927335401121\\
};

\addplot [very thick, color=mycolor1, mark=o, mark options={solid, mycolor1}]
  table[row sep=crcr]{%
5	0.639525965539224\\
10	0.531294220403874\\
15	0.498556023023754\\
20	0.484989522065803\\
25	0.465306203738568\\
30	0.455913071759532\\
};

\addplot [color=mycolor3, mark=o, mark options={solid, mycolor3}]
  table[row sep=crcr]{%
5	0.195233320626858\\
10	0.232501170435323\\
15	0.242712639204574\\
20	0.218182744877878\\
25	0.195560368762513\\
30	0.178697229920036\\
};

\addplot [color=mycolor4, mark=o, mark options={solid, mycolor4}]
  table[row sep=crcr]{%
5	0.19242878452998\\
10	0.0916200424411684\\
15	0.0662295294700556\\
20	0.0515741137267073\\
25	0.0426878850134108\\
30	0.0368890708463239\\
};

\addplot [very thick, color=mycolor2, dashed, mark=asterisk, mark options={solid, mycolor2}]
  table[row sep=crcr]{%
5	0.772507629636806\\
10	0.668894322193206\\
15	0.613603606986114\\
20	0.578780390306734\\
25	0.570723638289432\\
30	0.546701213328483\\
};

\addplot [very thick, color=mycolor1, dashed, mark=asterisk, mark options={solid, mycolor1}]
  table[row sep=crcr]{%
5	0.909967118803773\\
10	0.776258903582\\
15	0.691112357846226\\
20	0.678319646206471\\
25	0.626044577057183\\
30	0.592103722472688\\
};

\addplot [color=mycolor3, dashed, mark=asterisk, mark options={solid, mycolor3}]
  table[row sep=crcr]{%
5	0.539485968853007\\
10	0.594513948998464\\
15	0.603488590466242\\
20	0.592556386647332\\
25	0.575526502735339\\
30	0.555948049307371\\
};

\addplot [color=mycolor4, mark=asterisk, mark options={solid, mycolor4}]
  table[row sep=crcr]{%
5	0.175672031713911\\
10	0.0916200424411684\\
15	0.0662295294700556\\
20	0.0515741137267073\\
25	0.0426878850134108\\
30	0.0368890708463239\\
};
\coordinate (bot) at (rel axis cs:1,0);
\end{groupplot}

\path (top|-current bounding box.north)--
      coordinate(legendpos)
      (bot|-current bounding box.north);
\matrix[
    matrix of nodes,
    anchor=south,
    draw,
    inner sep=0.2em,
    draw,
    font = \normalsize,
  ]at([yshift=1.5ex]legendpos)
  {$\gamma_1=0.01$:&[5pt]
\ref{plots:NKD(a)}& NKD&[5pt]
\ref{plots:Sep(a)}& Unfold&[5pt]
\ref{plots:Flat(a)}& Flatten \cite{segarra2017network}&[5pt]
\ref{plots:Pgl(a)}& PGL \cite{kadambari2021product}\\
$\gamma_1=0.33$:&[5pt]
\ref{plots:NKD(b)}& NKD&[5pt]
\ref{plots:Sep(b)}& Unfold&[5pt]
\ref{plots:Flat(b)}& Flatten \cite{segarra2017network}&[5pt]
\ref{plots:Pgl(b)}& PGL \cite{kadambari2021product}\\
};
\end{tikzpicture}%
}\vspace{-.4cm}
\caption{\textbf{Topology Reconstruction.} $F_{1}$ score against (Left) sample size $S$ with $N =10$; (Right) graph size $N$ with $S=1000$.
}\vspace{-.4cm}
\label{fig:topology}
\end{figure}
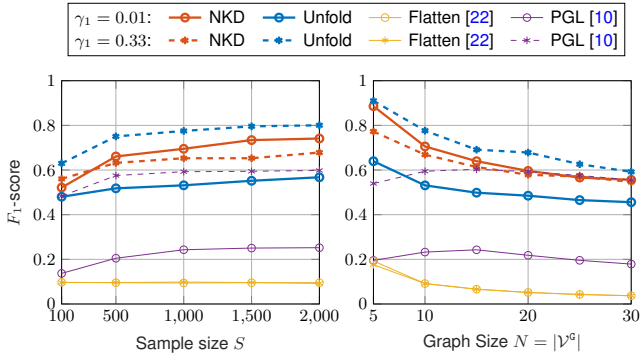

We consider synthetic data generated from \eqref{eq:gsp} with the graph filter given by $ \HAJ=e^{\tau \A^{\tt I}}$, where $\A^{\tt I}$ is parameterized by $\gamma_1 \in [0,\frac{1}{3}]$ as described. We also set $\tau = 1 / \max_i d_i$ with $d_i = \sum_{j=1}^N \A^{\tt I}_{ij}$. Excitation and noise signals satisfy ${\bm x}^{(s)} \sim {\cal N}( {\bm 0}, {\bm I} )$, ${\bm w}^{(s)} \sim {\cal N}( {\bm 0}, 0.01 {\bm I} )$.
Fig.~\ref{fig:topology} compares the performance of topology reconstruction in terms of the $F_1$ score against the sample size $S$ and graph size $N$. We assume known $\gamma_1$ and evaluate the $F_1$ scores by comparing the ground truth to $\widehat{\A}^{\tt I}$ reconstructed by \eqref{eq:gene_graph} using the estimated $\A^{\tt C}, \A^{\tt G}$.
We observe that `NKD' recovers the graph topology for a wide range of sample sizes and graph sizes, regardless of the coupling parameter $\gamma_1$. Meanwhile, `Unfold' and `PGL' are sensitive to $\gamma_1$ -- their performances are comparable to Algorithm~\ref{algo:prod_learn} when $\gamma_1 = \frac{1}{3}$, but it degrades noticeably when $\gamma_1 = 0.01$. In all cases, `Flatten' fails to estimate the graph topology.\vspace{.1cm}


\noindent \textbf{Example 2: Central Nodes Detection.}
The next example examines the performance of central node detection ({\bf P2}). We focus on detecting the central nodes of ${\cal G}^{\tt G}$ using \eqref{eq:centrali}, where ${\cal G}^{\tt G}$ is generated as a core-periphery graph \cite{core_per}. The node set $\mathcal{V}^{\tt G}=\{1,\ldots,N\}$ is partitioned into $\Vcore=\{1,\ldots,10\}$, $\Vpe=\mathcal{V}^{\tt G}\backslash\Vcore$. For every pair $(i,j) \in \mathcal{V}^{\tt G} \times \mathcal{V}^{\tt G}$, edges are assigned randomly with probability $1$ if $i,j \in \Vcore$; with probability $0.2$ if $i \in \Vcore, j \in \Vpe$; with probability $0.05$ if $i,j \in \Vpe$. Our goal is to detect the $10$ nodes in $\Vcore$ based on the observed multi-attribute graph signals. We focus on comparing the proposed `NKD' based Algorithm~\ref{algo:prod_learn} with `Unfold' which estimates $\widetilde{\V}^{\tt C}, \widetilde{\V}^{\tt G}$ from the node-wise unfolded graph signals. 

\begin{figure}[t]
\centering
\resizebox{0.995\linewidth}{!}{\sf \definecolor{mycolor1}{rgb}{0.00000,0.44700,0.74100}%
\definecolor{mycolor2}{rgb}{0.85000,0.32500,0.09800}%
\definecolor{mycolor3}{rgb}{0.92900,0.69400,0.12500}%
\definecolor{mycolor4}{rgb}{0.49400,0.18400,0.55600}%

\begin{tikzpicture}

\begin{groupplot}[group style={group name=myplot,group size=2 by 1}]
\nextgroupplot[%
width=2.5in,
height=2.25in,
x grid style={white!69.0196078431373!black},
xmajorgrids,
y grid style={white!69.0196078431373!black},
ymajorgrids,
xmode=log,
xmin=50,
xmax=20000,
 xminorticks=true,
xlabel style={font=\color{white!15!black}},
xlabel={Sample Size $S$},
ymin=0,
ymax=0.9,
ylabel style={font=\color{white!15!black}},
ylabel={Error Rate},
axis background/.style={fill=white},
legend style={legend cell align=left, align=left, draw=white!15!black}
]
\addplot [very thick, color=mycolor2, mark=o, mark options={solid, mycolor2}]
  table[row sep=crcr]{%
50	0.890000000000001\\
100	0.843\\
200	0.675\\
500	0.376\\
1000	0.144\\
2000	0.033\\
5000	0.001\\
8000	0\\
20000	0\\
};
\label{plots:NKD-0.01,inv}
\addplot [very thick, color=mycolor1, mark=o, mark options={solid, mycolor1}]
  table[row sep=crcr]{%
50	0.854\\
100	0.865\\
200	0.812\\
500	0.702\\
1000	0.547\\
2000	0.31\\
5000	0.0909999999999999\\
8000	0.045\\
20000	0.001\\
};
\label{plots:Separable-0.01,inv}
\addplot [very thick, color=mycolor3, mark=o, mark options={solid, mycolor3}]
  table[row sep=crcr]{%
50	0\\
100	0\\
200	0\\
500	0\\
1000	0\\
2000	0\\
5000	0\\
8000	0\\
20000	0\\
};
\label{plots:NKD-0.01,exp}

\addplot [very thick, color=mycolor4, mark=o, mark options={solid, mycolor4}]
  table[row sep=crcr]{%
50	0\\
100	0\\
200	0\\
500	0\\
1000	0\\
2000	0\\
5000	0\\
8000	0\\
20000	0\\
};
\label{plots:Separable-0.01,exp}

\addplot [very thick, color=mycolor2, dashed, mark=asterisk, mark options={solid, mycolor2}]
  table[row sep=crcr]{%
50	0.831\\
100	0.594\\
200	0.338\\
500	0.0949999999999999\\
1000	0.025\\
2000	0.003\\
5000	0\\
8000	0\\
20000	0\\
};
\label{plots:NKD-0.33,inv}

\addplot [very thick, color=mycolor1, dashed, mark=asterisk, mark options={solid, mycolor1}]
  table[row sep=crcr]{%
50	0.419\\
100	0.273\\
200	0.0989999999999999\\
500	0.016\\
1000	0.004\\
2000	0\\
5000	0\\
8000	0\\
20000	0\\
};
\label{plots:Separable-0.33,inv}

\addplot [very thick, color=mycolor3, dashed, mark=asterisk, mark options={solid, mycolor3}]
  table[row sep=crcr]{%
50	0\\
100	0\\
200	0\\
500	0\\
1000	0\\
2000	0\\
5000	0\\
8000	0\\
20000	0\\
};
\label{plots:NKD-0.33,exp}

\addplot [very thick, color=mycolor4, dashed, mark=asterisk, mark options={solid, mycolor4}]
  table[row sep=crcr]{%
50	0\\
100	0\\
200	0\\
500	0\\
1000	0\\
2000	0\\
5000	0\\
8000	0\\
20000	0\\
};
\label{plots:Separable-0.33,exp}
\coordinate (top) at (rel axis cs:0,1);

\nextgroupplot[%
width=2.5in,
height=2.25in,
x grid style={white!69.0196078431373!black},
xmajorgrids,
y grid style={white!69.0196078431373!black},
ymajorgrids,
xmin=20,
xmax=100,
xlabel style={font=\color{white!15!black}},
xlabel={Graph Size $N=|{\cal V}^{\tt G}|$},
ymin=0,
ymax=0.9,
ylabel style={font=\color{white!15!black}},
axis background/.style={fill=white},
legend style={legend cell align=left, align=left, draw=white!15!black}
]
\addplot [very thick, color=mycolor2, mark=o, mark options={solid, mycolor2}]
  table[row sep=crcr]{%
20	0.03\\
30	0.055\\
40	0.071\\
50	0.084\\
60	0.0999999999999999\\
80	0.123\\
100	0.116\\
};

\addplot [very thick, color=mycolor1, mark=o, mark options={solid, mycolor1}]
  table[row sep=crcr]{%
20	0.0869999999999999\\
30	0.161\\
40	0.247\\
50	0.3\\
60	0.359\\
80	0.502\\
100	0.589\\
};

\addplot [very thick, color=mycolor3, mark=o, mark options={solid, mycolor3}]
  table[row sep=crcr]{%
20	0\\
30	0.001\\
40	0\\
50	0\\
60	0\\
80	0\\
100	0\\
};

\addplot [very thick, color=mycolor4, mark=o, mark options={solid, mycolor4}]
  table[row sep=crcr]{%
20	0\\
30	0.001\\
40	0\\
50	0\\
60	0\\
80	0\\
100	0\\
};

\addplot [very thick, color=mycolor2, dashed, mark=asterisk, mark options={solid, mycolor2}]
  table[row sep=crcr]{%
20	0.006\\
30	0.01\\
40	0.012\\
50	0.016\\
60	0.01\\
80	0.011\\
100	0.014\\
};

\addplot [very thick, color=mycolor1, dashed, mark=asterisk, mark options={solid, mycolor1}]
  table[row sep=crcr]{%
20	0\\
30	0.005\\
40	0.002\\
50	0.002\\
60	0.002\\
80	0.002\\
100	0.001\\
};

\addplot [very thick, color=mycolor3, dashed, mark=asterisk, mark options={solid, mycolor3}]
  table[row sep=crcr]{%
20	0\\
30	0.001\\
40	0\\
50	0\\
60	0\\
80	0\\
100	0\\
};

\addplot [very thick, color=mycolor4, dashed, mark=asterisk, mark options={solid, mycolor4}]
  table[row sep=crcr]{%
20	0\\
30	0.001\\
40	0\\
50	0\\
60	0\\
80	0\\
100	0\\
};

\coordinate (bot) at (rel axis cs:1,0);
\end{groupplot}

\path (top|-current bounding box.north)--
      coordinate(legendpos)
      (bot|-current bounding box.north);
\matrix[
    matrix of nodes,
    anchor=south,
    draw,
    inner sep=0.2em,
    draw,
    font = \normalsize,
  ]at([yshift=1.5ex]legendpos)
  {$\gamma_1=0.01$:&[1pt]
\ref{plots:NKD-0.01,inv}& NKD ${\cal H}^{\sf inv}$&[.5pt]
\ref{plots:Separable-0.01,inv}& Unfold ${\cal H}^{\sf inv}$ &[.5pt]
\ref{plots:NKD-0.01,exp}& NKD ${\cal H}^{\sf exp}$ &[.5pt]
\ref{plots:Separable-0.01,exp}& Unfold ${\cal H}^{\sf exp}$ \\
$\gamma_1=0.33$:&[1pt]
\ref{plots:NKD-0.33,inv}& NKD ${\cal H}^{\sf inv}$&[.5pt]
\ref{plots:Separable-0.33,inv}& Unfold ${\cal H}^{\sf inv}$ &[.5pt]
\ref{plots:NKD-0.33,exp}& NKD ${\cal H}^{\sf exp}$&[.5pt]
\ref{plots:Separable-0.33,exp}& Unfold ${\cal H}^{\sf exp}$\\
};
\end{tikzpicture}%
}\vspace{-.4cm}
\caption{\textbf{Central Nodes Detection.} Error rate against (Left) sample size $S$ with $N=80$; (Right) graph size $N$ with $S=5MN$.
}\vspace{-.2cm}
\label{fig:cen_de}
\end{figure}
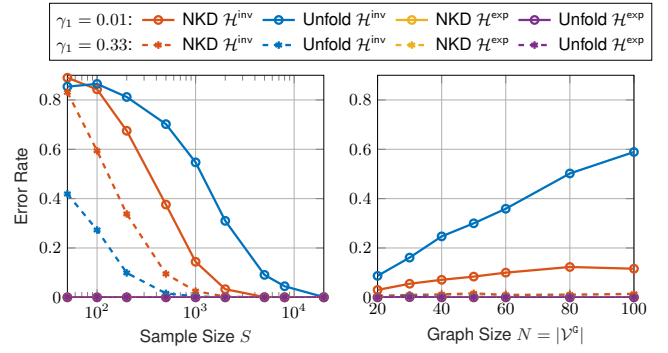

We consider synthetic data modeled after \eqref{eq:gsp} with two graph filters: (a) ${\cal H}^{\sf inv}( \A^{\tt C}, \A^{\tt G} ) =(\I- \tau_1 \A^{\tt I})^{-1}$, (b) ${\cal H}^{\sf exp}( \A^{\tt C}, \A^{\tt G} ) = e^{\tau_2 \A^{\tt I}}$ with $\tau_1 = 1/\max_i d_i$, $\tau_2 = 10\tau_1$. We set ${\bm x}^{(s)} \sim {\cal N}( {\bm 0}, {\bm I} )$, ${\bm w}^{(s)} \sim {\cal N}( {\bm 0}, 0.01 {\bm I} )$.
Fig.~\ref{fig:cen_de} compares the error rate in identifying the central nodes $\Vcore$ as top-10 central nodes \eqref{eq:centrali}. As seen, `NKD' delivers lower error rate with smaller number of samples for the strong coupling scenario ($\gamma_1 = 0.01$). The result corroborates with Proposition~\ref{prop:sep} as the covariance with node/layer-wise unfolded signals have close (or repeated) eigenvalues, significantly increasing the sample complexity in eigenvector estimation.\vspace{.1cm}

\begin{figure}[t]
\centering
{\resizebox{0.47\linewidth}{!}{\sf \includegraphics{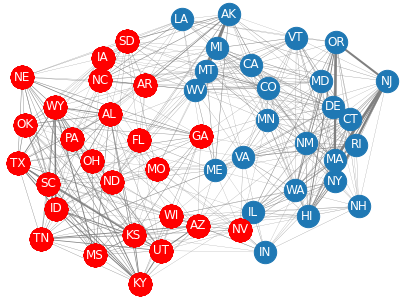}}
}
{\resizebox{0.45\linewidth}{!}{\sf \includegraphics{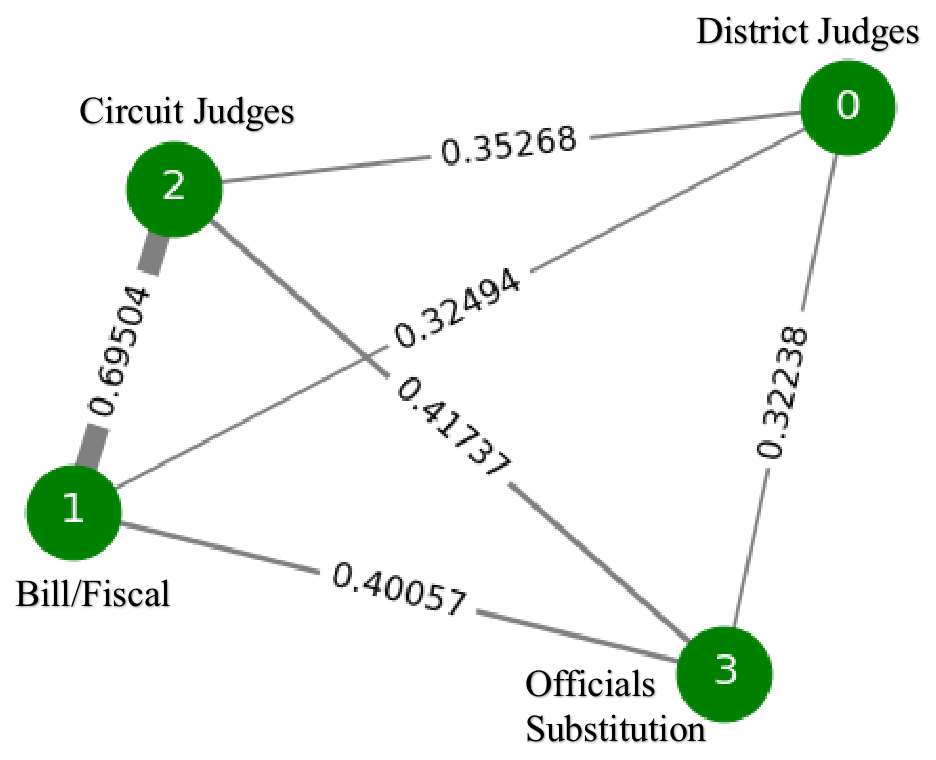}}
}%
\vspace{-.25cm}
\caption{\textbf{US Senate Rollcalls Data.} (Left) Estimated Senate topology ${\cal G}^{\tt G}$ color coded by the two communities found by spectral clustering on ${\bm A}^{\tt G}$.
(Right) Estimated coupling graph of the topics of rollcalls.
}\vspace{-.55cm}
\label{fig:rollcall}
\end{figure}

\noindent \textbf{Real Data: US Senate Roll Calls.}
We apply Algorithm~\ref{algo:prod_learn} to reconstruct the topology ({\bf P1}) from the 113th-116th US Senate rollcalls [available: \url{https://voteview.com/data}]. 
We first apply NMF \cite{arora2012learning} to perform topic modeling
on rollcalls' descriptions to get $M=4$ topics, we then model and infer a product graph with $M=4$ layers and $N=50$ nodes, where each node corresponds to the Senators from a state. 
Fig.~\ref{fig:rollcall} shows the learnt graph topology from $S=101$ samples through applying Algorithm~\ref{algo:prod_learn}. The estimated topology is reasonable: the state graph ${\cal G}^{\tt G}$ identifies the Republican and Democratic states as two clusters, and the coupling graph ${\cal G}^{\tt C}$ shows a strong connection between topics on `circuit judges' and `fiscal'.\vspace{.1cm}

\noindent \textbf{Conclusions.} This paper considered the problem of learning from multi-attribute graph signals by proposing a general product graph filter model and developing its inference algorithms. 
Future works include analyzing the sampling complexity of Algorithm~\ref{algo:prod_learn} and studying the effects of layer-wise/node-wise unfolding of graph signals on sampling complexity or identifiability.

\newpage
\bibliographystyle{IEEEtran}
\bibliography{ref}

\iftrue
\newpage 
\appendix 
\section{Proof of Proposition~2} \label{app:sep}
We first show the identity for the node-wise covariance ${\bm C}_y^{\sf node}$. Define the following selection matrix for $m=1,\ldots,M$, 
\[
{\bm C}_m = \big[ {\bm 0} ~\cdots~ {\bm I}_{N} ~\cdots~ {\bm 0} \big] = {\bm e}_m^\top \otimes {\bm I}_N \in \RR^{N \times NM},
\]
where ${\bm e}_m$ is the $m$th canonical basis vector for $\RR^M$. 
We observe that 
\beq \label{eq:YYT} \textstyle 
{\bm Y}^{(s)} ( {\bm Y}^{(s)} )^\top = \sum_{m=1}^M {\bm C}_m {\bm y}^{(s)} ( {\bm y}^{(s)} )^\top {\bm C}_m^\top 
\eeq 
Under the assumptions of the proposition, we observe that for any $m=1,\ldots,M$,
\[
\begin{split}
& \EE[ {\bm C}_m {\bm y}^{(s)} ( {\bm y}^{(s)} )^\top {\bm C}_m^\top ] \\
& = \sum_{i=1}^M \sum_{j=1}^N |h(\lambda_i^{\tt C}, \lambda_j^{\tt G})|^2 {\bm C}_m ( {\bm v}_i^{\tt C} \otimes {\bm v}_j^{\tt G}) ( {\bm v}_i^{\tt C} \otimes {\bm v}_j^{\tt G})^\top {\bm C}_m^\top \\
& = \sum_{j=1}^N {\bm v}_j^{\tt G} ({\bm v}_j^{\tt G})^\top \sum_{i=1}^M |h(\lambda_i^{\tt C}, \lambda_j^{\tt G})|^2 |v_{i,m}^{\tt C}|^2  
\end{split}
\]
Substituting back into \eqref{eq:YYT} yields
\[ 
\begin{split} 
{\bm C}_y^{\sf node} & = \sum_{j=1}^N {\bm v}_j^{\tt G} ({\bm v}_j^{\tt G})^\top \sum_{i=1}^M |h(\lambda_i^{\tt C}, \lambda_j^{\tt G})|^2 \sum_{m=1}^M |v_{i,m}^{\tt C}|^2 \\
& = \sum_{j=1}^N {\bm v}_j^{\tt G} ({\bm v}_j^{\tt G})^\top \sum_{i=1}^M |h(\lambda_i^{\tt C}, \lambda_j^{\tt G})|^2. 
\end{split} 
\]
Next, we consider the layer-wise covariance ${\bm C}_y^{\sf layer}$. Similarly, we define the selection matrix for $n=1,\ldots,N$, 
\[
{\bm D}_n = {\bm I}_M \otimes {\bm e}_n^\top \in \RR^{M \times NM},
\] 
where ${\bm e}_n$ in the above is the $n$th canonical basis vector for $\RR^N$. We observe that 
\beq \label{eq:YTY}
( {\bm Y}^{(s)} )^\top {\bm Y}^{(s)} = \sum_{n=1}^N {\bm D}_n {\bm y}^{(s)} ( {\bm y}^{(s)} )^\top {\bm D}_n^\top
\eeq 
For any $n = 1,\ldots,N$, it holds
\[
\begin{split}
& \EE[ {\bm D}_n {\bm y}^{(s)} ( {\bm y}^{(s)} )^\top {\bm D}_n^\top ] \\
& = \sum_{i=1}^M \sum_{j=1}^N |h(\lambda_i^{\tt C}, \lambda_j^{\tt G})|^2 {\bm D}_n ( {\bm v}_i^{\tt C} \otimes {\bm v}_j^{\tt G}) ( {\bm v}_i^{\tt C} \otimes {\bm v}_j^{\tt G})^\top {\bm D}_n^\top \\
& = \sum_{i=1}^M {\bm v}_i^{\tt C} ({\bm v}_i^{\tt C})^\top \sum_{j=1}^N |h(\lambda_i^{\tt C}, \lambda_j^{\tt G})|^2 |v_{j,n}^{\tt G}|^2  
\end{split}
\]
Substituting back into \eqref{eq:YTY} yields
\[ 
\begin{split} 
{\bm C}_y^{\sf layer} & = \sum_{i=1}^M {\bm v}_i^{\tt C} ({\bm v}_i^{\tt C})^\top \sum_{j=1}^N |h(\lambda_i^{\tt C}, \lambda_j^{\tt G})|^2 \sum_{n=1}^N |v_{j,n}^{\tt G}|^2 \\
& = \sum_{i=1}^M {\bm v}_i^{\tt C} ({\bm v}_i^{\tt C})^\top \sum_{j=1}^N |h(\lambda_i^{\tt C}, \lambda_j^{\tt G})|^2.
\end{split} 
\]
This concludes the proof.
\fi

\end{document}